\documentclass[journal]{vgtc}                

\ifpdf
  \pdfoutput=1\relax                   
  \usepackage{graphicx}                
  \DeclareGraphicsExtensions{.pdf,.png,.jpg,.jpeg} 
\else
  \usepackage{graphicx}                
  \DeclareGraphicsExtensions{.eps}     
\fi%

\graphicspath{{figures/}{pictures/}{images/}{./}} 

\usepackage{microtype}                 
\PassOptionsToPackage{warn}{textcomp}  
\usepackage{textcomp}                  
\usepackage{mathptmx}                  
\usepackage{times}                     
\usepackage{cite}                      
\usepackage{tabu}                      
\usepackage{booktabs}                  
\usepackage{amsmath, xparse}
\usepackage[dvipsnames]{xcolor}

\usepackage{subfigure}
\usepackage[normalem]{ulem}
\usepackage{algorithmic}
\usepackage[ruled, noline, linesnumbered]{algorithm2e}

\onlineid{1413}

\vgtccategory{Research}

\newcommand{\mytodoblack}[1]{\textcolor{black}{{\sf}#1}}

\definecolor{orange}{rgb}{0.1,0.48,0.0}

\newcommand{\remove}[1]{}

\newcommand{\add}[1]{\mytodoblack{#1}}

\vgtcpapertype{algorithm/technique}

\title{HiTailor: Interactive Transformation and Visualization for 
\\
Hierarchical Tabular Data}

\author{Guozheng Li, Runfei Li, Zicheng Wang, Chi Harold Liu, Min Lu, and Guoren Wang}
\authorfooter{
\item
 Guozheng Li, Runfei Li, Zicheng Wang, Chi Harold Liu and Guoren Wang are with Beijing Institute of Technology. E-mail: \{guozheng.li, chiliu\}@bit.edu.cn. Runfei Li and Zicheng Wang contribute equally to this work. Chi Harold Liu is the corresponding author.  
\item
 Min Lu is with Shenzhen University. E-mail: lumin.vis@gmail.com
}

\shortauthortitle{Biv \MakeLowercase{\textit{et al.}}: Global Illumination for Fun and Profit}

\abstract{%
Tabular visualization techniques integrate visual representations with tabular data to avoid additional cognitive load caused by splitting users' attention. 
However, \add{most of the} existing studies focus on simple flat tables instead of hierarchical tables, whose complex structure limits the expressiveness of visualization results and affects users' efficiency in visualization construction.
We present HiTailor, a technique for presenting and exploring hierarchical tables. 
HiTailor constructs an abstract model, which defines row/column headings as biclustering and hierarchical structures. 
Based on our abstract model, we identify three pairs of operators, Swap/Transpose, ToStacked/ToLinear, Fold/Unfold, for transformations of hierarchical tables to support users' comprehensive explorations.
After transformation, users can specify a cell or block of interest in hierarchical tables as a TableUnit for visualization, and HiTailor recommends other related TableUnits according to the abstract model using different mechanisms. 
We demonstrate the usability of the HiTailor system through a comparative study and a case study with domain experts, showing that HiTailor can present and explore hierarchical tables from different viewpoints. 
\add{HiTailor is available at \href{https://github.com/bitvis2021/HiTailor}{https://github.com/bitvis2021/HiTailor}.}

} 


\CCScatlist{ 
 \CCScat{K.6.1}{Management of Computing and Information Systems}%
{Project and People Management}{Life Cycle};
 \CCScat{K.7.m}{The Computing Profession}{Miscellaneous}{Ethics}
}

\teaser{
  \centering
  \includegraphics[width=0.95\linewidth]{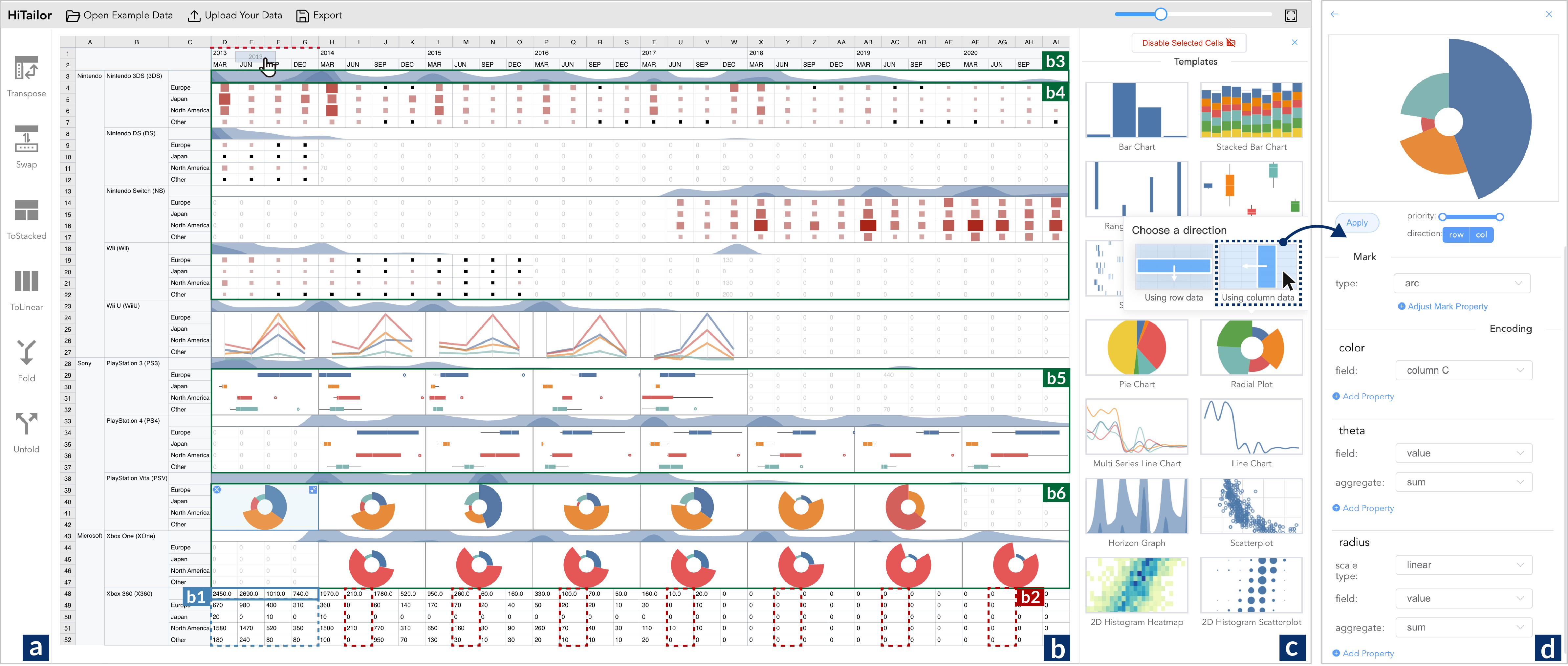}
  \caption{ The user interface of the HiTailor prototype system. (a) Transformation operator panel; (b) Tabular visualization panel; (c) Visualization template panel; (d) Visualization configuration panel.
  }
\label{fig:teaser}
}

\vgtcinsertpkg

\begin{document}



\firstsection{Introduction}

\maketitle
Tabular data has been a critical data management approach and is widely adopted by many application domains, including scientists, financial practitioners, and policy-makers~\cite{2018-Expandable-Dou, 2013-Automatic-Chen}. 
Tabular visualizations (TVs) either encode a data item within a cell into a single visual element~\cite{2014-Revisiting-charles, 1994-table-rao, 2013-Lineup-Gratzl, 2016-heatmap-gu} or integrate summary visualizations for a block containing a rectangular group of continuous cells with tabular data to reveal its overview and patterns~\cite{2011-visbrick, 2019-taggle-iv}. 
TVs retain a tabular layout, and the integration of data and visual representations could avoid additional cognitive load caused by splitting the users’ attention~\cite{2006-AyersSweller, 2005-implications-sweller, 1998-cognitive-sweller, 2006-integrating-ginns}. 

\add{Most of the existing studies} on TVs focus on simple flat tables~\cite{2019-taggle-iv, 2014-Revisiting-charles} but neglect hierarchical tables, whose headings exhibit multi-level structures.
Hierarchical tables are widely used, especially in statistical reports and research papers. 
A research study~\cite{2013-Automatic-Chen} about web spreadsheet corpus shows that hierarchical tables account for 32.5\% of the tabular dataset.
The hierarchical structure~\cite{2020-barcodetree-li, 2020-gotree-li} in column or row headings leads to better capability of efficient data management~\cite{2021-TabularNet-Du}. 
However, hierarchical headings also build incompatible logical and positional relationships between data items.
More specifically, cells in the tabular data with the same relative positions might have different relationships. 
For example, Fig.~\ref{fig:teaser} shows the sales of different types of game consoles by Nintendo, Sony and Microsoft in different regions from 2013 to 2020.
Each cell in the top row of $b1$ is the sum of the corresponding values in the rows below (with a dashed border in blue), while the inner cells of these rows below are comparable attributes.
In addition, some closely related items in the hierarchical tabular data are not adjacent to each other. For example, the cells in $b2$ (with a dashed border in red) describe the sales of Xbox 360 in June for different years but are separated from each other in the tabular data.
The inconsistency of hierarchical tabular data poses challenges using TVs. 
First, to create high-quality visual representations, users need to specify several discrete blocks of tabular data and transform them into visualizations repeatedly, making the whole authoring process tedious and time-consuming. 
Second, to allow exploring tabular data from different viewpoints using TVs, users need to change the positions of cells within tabular data on demand. 
Specifically, they need to place the associated cells in a contiguous block, enabling the integration of visualization results with tabular data.

We present HiTailor (\textbf{Hi}erarchical \textbf{Ta}ble \textbf{Il}lustrat\textbf{or}), a technique to support interactive transformation and visualization for hierarchical tables. 
HiTailor is based on the abstract model of hierarchical tables by parsing the column and row headings as hierarchical or biclustering~\cite{2018-BiDots-Zhao} structures. 
We identify three pairs of operators for hierarchical table transformations, including Swap/Transpose, ToStacked/ToLinear, and Fold/Unfold. 
Different operators for table transformations are unified together based on the abstract model, allowing users to transform hierarchical tables continuously.

Tabular data transformations reorganize hierarchical tables and place cells together that need to be aggregated and visualized.
From the transformation results, users can select a single cell or a rectangular group of contiguous cells (\textit{i.e.}, block) within the tabular data as a \textit{TableUnit} and visualize using different techniques, including unit visualization and summary visualization based on Vega-Lite. 
After users specify a TableUnit within hierarchical tables, HiTailor recommends corresponding blocks with different priorities according to their relative positions with users' selection in the abstract model.  
We design different TableUnit recommendation mechanisms to meet users' various requirements for visualizing hierarchical tables, avoiding the repeated and tedious manual specifications and improving the efficiency of tabular visualization construction.
We implement the HiTailor prototype system to support both exploration and presentation of hierarchical tables.

We validate HiTailor through a use case and a user study.
First, we demonstrate the usability of HiTailor with a use case in real-world application scenarios, and the results show that HiTailor enables users to explore and present the hierarchical table from different viewpoints. 
\add{Second, we validate the effectiveness of the HiTailor prototype system through a comparative study with Tableau, and the results show that HiTailor is more efficient in building an overview for the hierarchical table as well as analyzing data under different levels of headings.}


In summary, the main contributions of this paper are as follows.
\add{First, we construct an abstract model that identifies hierarchies and biclusters from the headings of hierarchical tables for transformations and visualizations.}
\add{Second, we define TableUnit and present recommendation mechanisms based on the abstract model for improving the efficiency of constructing hierarchical table visualizations}.
Third, we propose the HiTailor prototype system for presenting and exploring hierarchical tables with validating the utilities and effectiveness. 

\add{The layout of the rest of this paper is as follows.
In Section 2, we survey existing approaches to the transformation and visualization of tabular data. 
Next, we describe in Section 3 the detailed techniques of HiTailor. 
We then present the design and implementation of the HiTailor prototype system in Section 4 before conducting a case study (Section 5) and a user study (Section 6) to validate the effectiveness of our method. 
Finally, we conclude with some discussion and directions for future work.
}

\vspace{-0.1cm}
\section{Related Work}
A large number of tools and studies exist for the visualization and analysis of tabular data. Our literature review emphasizes two critical aspects, tabular data transformation and visualization. 

\subsection{Tabular Data Transformation}

Most studies about tabular data transformation rearrange the tables with arbitrary forms into a relational schema. These transformations remove the hierarchical structures and redundant information from tables and enable users to import the transformation results into databases or other data tools. 
We divide these studies into two categories, \textit{non-automatic} and \textit{automatic}, based on whether human involvement is required. 
\add{In addition, we also compare HiTailor with Pivot Table and Tableau, which support the tabular data analysis through data transformations.}
Among the non-automatic transformation methods, some work requires users to define transformation rules explicitly by low-level programming.
TranSheet\cite{2011-spreadsheet-hung} provides a spreadsheet-like formula mapping language that allows users to perform five value-related and ten structure-related transformations to specify mappings between source spreadsheet data and a structured form.
Existing studies~\cite{2019-TabbyXL-Alexey, 2022-rigel-chen} also show new possibilities in understanding different data transformations, which allows users to define functional and structural relationships of tables using domain-specific language, giving interpretations simultaneously.
These approaches are highly flexible, but learning new syntax about data transformation is difficult and time-consuming for users. 
In addition to the programming techniques, some work allows users to transform tabular data interactively. 
Potter's Wheel~\cite{2001-PotterWheel-Raman} provides an interactive user interface based on a declarative syntax and allows users to perform their desired transformations by clicking on different buttons.
Furthermore, Wrangler~\cite{2011-Wrangler-Sean} extends the transformation language with a recommendation engine to generate regular expressions, inferring subsequent transforming operations based on user interactions. 
However, the above studies focus on simple flat tables instead of hierarchical tables.
Xtable~\cite{1996-xtable-wang} builds an abstract model for hierarchical tables and provides an interactive environment for editing their logical structures.
\add{Nevertheless, the model is not designed for tabular data transformations, because it only builds the relationships between headings and entries but ignores the relationships between labels in headings.}
Some studies explore automatic techniques to improve the efficiency of tabular data transformations.
We divide these techniques into three categories, \textit{relational schema extraction}, \textit{example-oriented transformation}, and \textit{machine learning based methods}. 

The techniques based on relational schema extraction extract the relational schema from original tables explicitly and transform them into canonical relational tables automatically. These studies explore different heuristic extraction algorithms based on domain knowledge.
The \textit{CELLS} system~\cite{2015-table-shigarov} allows users to acquire the relational schema from hierarchical tables using assumptions, including general assumptions about relationships of table cells and special assumptions about spatial, style and natural language features. 
Furthermore, Su et al.~\cite{2017-Transforming-Su} model the structure of hierarchical tables and develop a method to generate a relational schema based on table boundary features and layout features.

Unlike the techniques for relational schema extraction, example-oriented transformations do not explicitly define the relational schema. Users provide the initial and final states in pairs as an example and these techniques generate a program automatically to implement the same transformations as described in the example pairs.
\add{ProgFromEx~\cite{2011-Spreadsheet-Harris} is an algorithm that uses filter programs and associative programs to infer the desired transformation according to example pairs given by users.}
Based on ProgFromEx, a DAG-based algorithm~\cite{2012-Spreadsheet-Gulwani} uses a \emph{generate} and \emph{intersect} strategy and defines ranking rules to infer the final transformation programs.
To reduce the difficulty of building example pairs, Flashrelate~\cite{2015-FlashRelate-Barowy} requires only some tuples as examples, including positive examples representing the desired transformations and negative examples representing the incorrect transformations. 

With the generation of tabular data and the development of machine learning methods, many studies explore more intelligent techniques for tabular data transformation.  
Chen et al.~\cite{2013-Automatic-Chen} use conditional random fields to identify the hierarchical structure of table headers. This method focuses on manually-engineered stylistic and formatting features of tabular cells or row/columns but ignore the spatial information, which shows how adjacent cells are organized.
Dong et al.~\cite{2019-semantic-dong} choose the multi-task learning architecture for spreadsheets' structure extraction. The method comprises two steps: extracting cell features and modelling table structures, while TabularNet~\cite{2021-TabularNet-Du} focuses on the second step to obtain a better result using a neural network structure.

\add{The above automatic techniques regard a table as a data format. They are mainly used to formalize the original structure to import tables into other data analysis tools. These techniques require a specific tabular structure as the transformation objective.
However, HiTailor presents a table as an interface for analysis, and users do not have a clear target for transformations at the beginning of their explorations.}

\add{Meanwhile, several tools for exploring tabular data exist, and the most typical ones are Pivot Table and Tableau. 
We compare HiTailor with these two systems from the following three perspectives.
First, HiTailor can directly take hierarchical tables as its input and recognize the structures from the headings of hierarchical tables. However, the other tools can only start from flat tables and require users to reorganize the original data as hierarchical tables interactively. 
Second, the transformations supported in HiTailor are designed to facilitate tabular visualization creation further. By changing the relative position of data items and adjusting the aspect ratio of the block where data items are arranged, the transformations enable users to create different visualizations using HiTailor. However, transformations in Pivot Table and Tableau are mainly used for the computation and processing of original tabular data, including sum and average.
Third, HiTailor allows users to transform the original table through intuitive direct manipulations, which helps users understand the results of transformations of tabular data. In comparison, Pivot Table and Tableau transform tabular data through shelf-based interactions, requiring users to place variables in rows or columns to organize table headings.}


\begin{figure*}[tb]
 \centering  
   \includegraphics[width=0.9\textwidth]{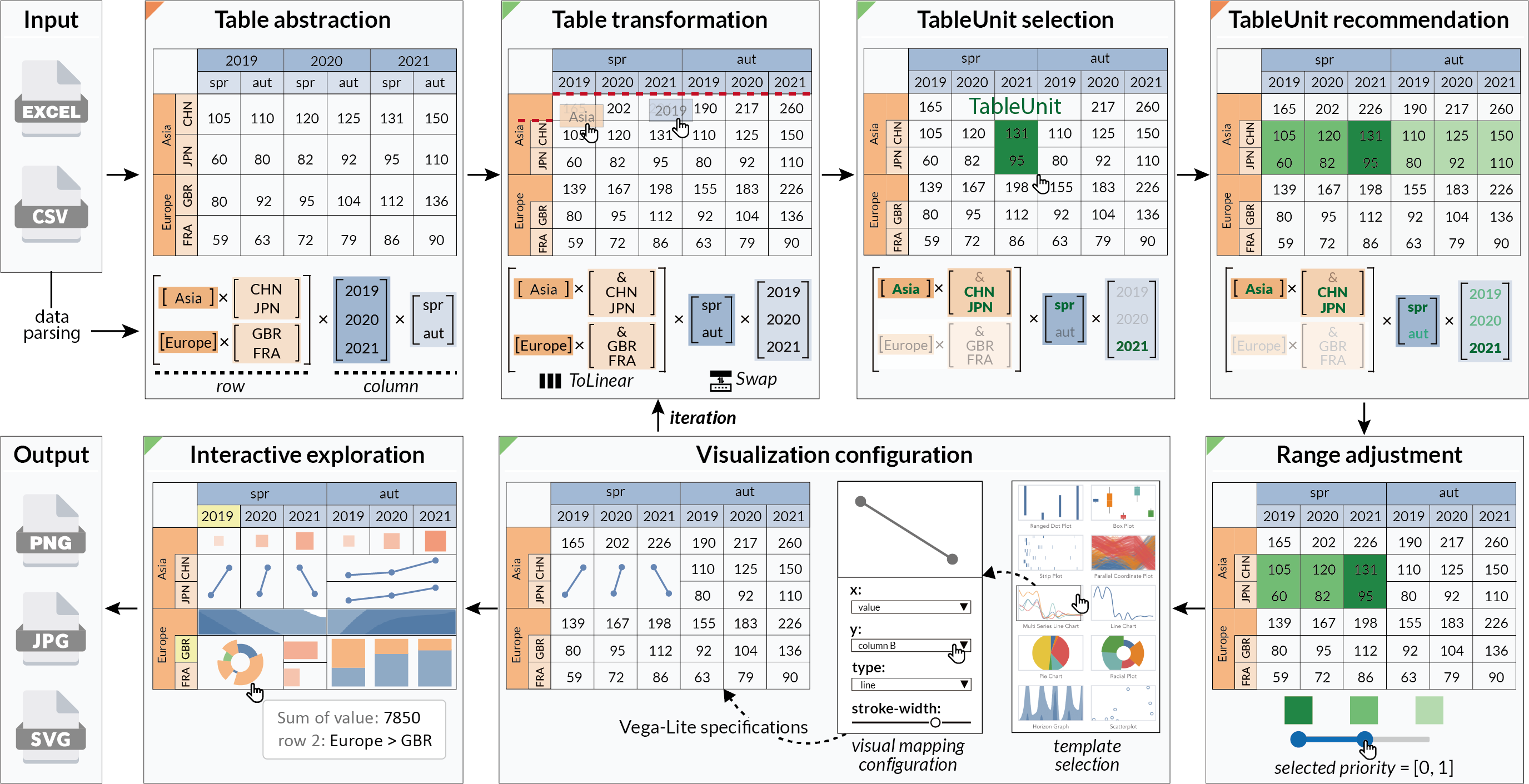}
 \caption{The pipeline of hierarchical table visualization with HiTailor. The visualization construction starts from the original tabular data. HiTailor constructs an abstract model by parsing the column and row headings of tabular data. Then users can transform the hierarchical tabular data by direct manipulation and select a specific \emph{TableUnit}. HiTailor recommends other related \emph{TableUnits} with different priorities. By setting the priority range, users can determine a series of targeted \emph{TableUnits}. For the visualization of \emph{TableUnit}, users select a visualization template from the gallery and adjust the visual mapping configurations. The visualization authoring is an iterative process based on the abstract model. Users can interactively explore the hierarchical table from the visualization results and export the results as images. In particular, the procedures with green marks require users' interaction, and the others with orange marks indicate automatic processes.}
 \label{fig:pipelineX}
\end{figure*}     

\vspace{-0.1cm}
\subsection{Tabular Data Visualization}
\label{sec:tabular-data-visualization}

The literature review in this section classifies research work from two aspects, visualization techniques and visualization targets.
Existing work~\cite{2019-taggle-iv} distinguishes three types of tabular data visualizations, overview, projection and tabular techniques. 
We mainly discuss the tabular techniques and hybrid techniques that combine overview and tabular approaches. 
The tabular techniques retain item positions across columns and encode the data within cells. 
The hybrid techniques preserve the relative positions between the data subset while the data items within the subsets are visualized using an overview.
We further divide these techniques into two categories according to the visualization targets, the first category focuses on a single table, and the second category builds the correlations between multiple tables.




The tabular techniques for a single table emphasize the values of individual items. 
Table Lens~\cite{1994-table-rao} explores large tabular data using a \emph{focus}+\emph{context} mechanism to encode cell values based on the data types of different columns. 
Furthermore, FOCUS~\cite{1996-focus-spenke} allows users to merge identical adjacent values to increase readability.
\add{Based on the matrix analysis approach for tabular data proposed by Jacques Bertin~\cite{1973-semiologie-bickmore}, CHART~\cite{1977-interactive-benson} and reorderable matrix~\cite{2005-constructing-harri, 2016-matrix-behrisch, 2004-visualising-dwyer, 2022-Simultaneous-Beusekom} embed visual elements in each cell and reorder matrices to create spatial associations between similar cells.}
Moreover, Bertifier~\cite{2014-Revisiting-charles} significantly increases functionality and interactivity, facilitating users to divide and annotate matrices into meaningful groups. 
Some work also represents a table matrix as a heat map, which encodes tabular data into colors, such as ComplexHeatmaps~\cite{2016-heatmap-gu}.
The above visualization tools focus on the deterministic tabular data, while Fuzzy Spreadsheet~\cite{2021-fuzzy-dhanoa} helps users track and explore uncertainty information, such as likelihood and probability distributions, to facilitate assumption analysis. 

The hybrid techniques aggregate multiple cells of tabular data to reveal patterns across attributes. ValueCharts~\cite{2004-ValueCharts-Carenini, 2006-integrated-bautista} addresses the multi-attribute ranking problem by visualizing each column according to a series of user-assigned weights using stacked bar chart visualizations. LineUp~\cite{2013-Lineup-Gratzl} establishes connections between stacked bars and integrates them into a slope graph, facilitating users to compare multiple rankings of the same set of items. 
Unlike the above studies, Podium~\cite{2017-podium-wall} collects users' preferences of data and inverts the appropriate attribute weights. Similar to HiTailor, Taggle~\cite{2019-taggle-iv} allows users to integrate visualizations for the aggregated cells within tabular data. 
However, it does not consider the logical structure of hierarchical tables. 

Other studies that focus on multiple tables try to build connections between them. 
For example, Domino~\cite{2014-Domino-Gratzl} chooses appropriate visualizations for each table within a dataset, connecting multiple subsets in various ways to show the relationships between them. 
TACO~\cite{2018-TACO-Niederer} focuses on the visual comparison of two different tables, which allows users to compare values and structures between multiple versions of the same table at different levels of granularity in an interactive way.

\add{However, existing studies about tabular visualizations always apply the visual encoding directly to the entries of same row and column, but does not focus on logical structure of a table with hierarchical headings.}

\vspace{-0.1cm}
\section{The Design of HiTailor}
\label{sec:hitailor-design}

This section introduces the detailed techniques of HiTailor, which comprises four parts, abstract model, hierarchical table transformation, TableUnit recommendation, and TableUnit visualization. In particular, the abstract model serves as the basis of the following three parts.  
Fig.~\ref{fig:pipelineX} presents the overall architecture of HiTailor.

\subsection{Abstract Model}
\label{sec:abstract-model}
We summarize the abstract model based on the hierarchical tabular data in real-world applications. We collect over three hundred hierarchical tables from the Bureau of Economic Analysis\footnote{www.bea.gov} and the State Statistical Bureau\footnote{www.stats.gov.cn/english}. We also utilize the tabular datasets from existing studies, including EUSES~\cite{2005-euses-fisher}, Fuse~\cite{2015-fuse-barik} and VEnron~\cite{2016-venron-dou}, which are three of the most widely-used corpora for tabular data research~\cite{2018-Expandable-Dou}.

\begin{figure}[tb] 
 \centering
 \includegraphics[width=0.87\columnwidth]{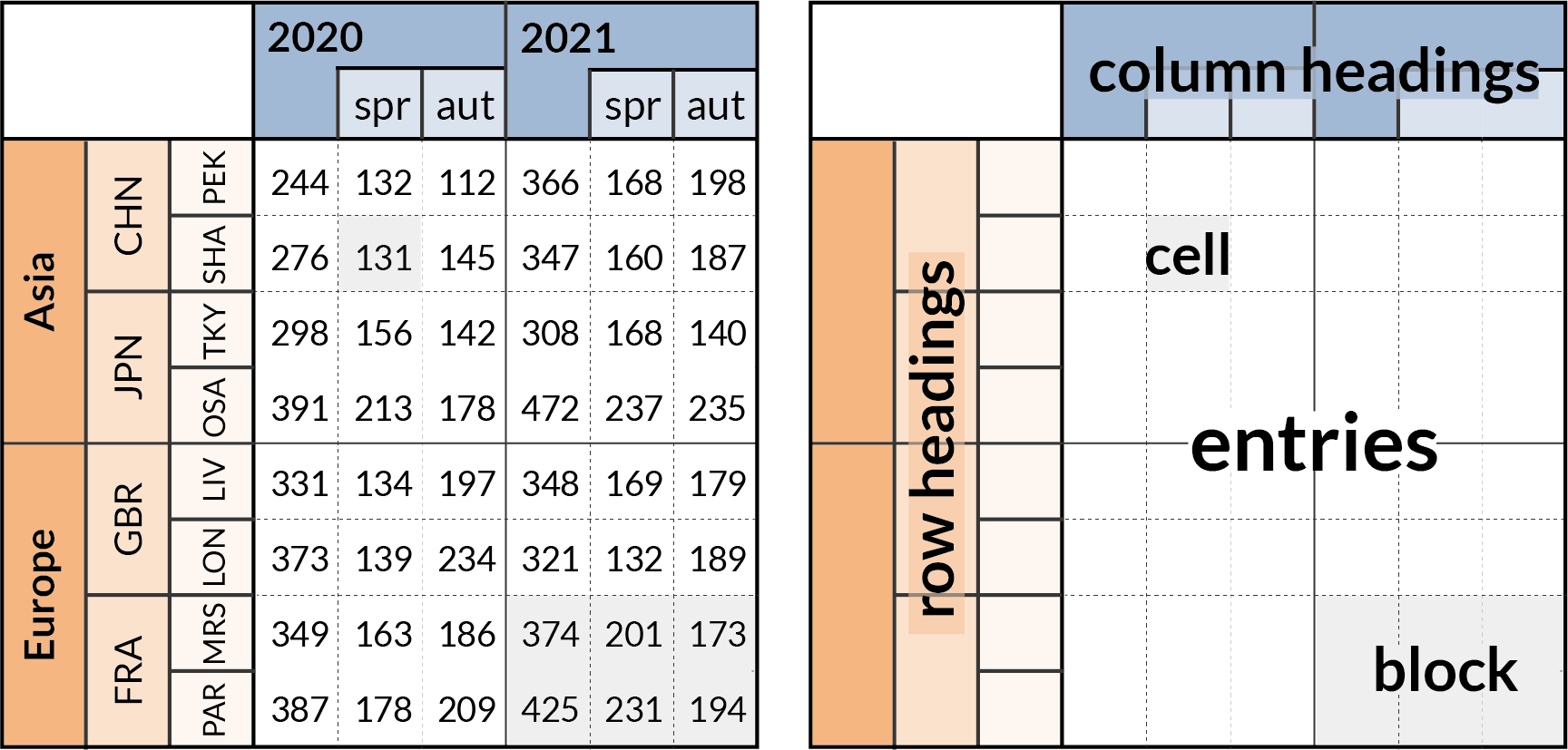}
 \caption{The left figure shows an example of a hierarchical table. The right figure marks primary components of the hierarchical table: row headings, column headings and entries. The cell indicates a single entry and the block indicates a continuous rectangular area consisting of multiple entries.}
 \label{fig:hierarchical-table-composition}
\end{figure}

A table is a collection of interrelated items, which can be divided into two groups: entries and labels~\cite{1996-xtable-wang}.
\emph{Entries} are the basic data items displayed in a table. 
\emph{Labels} are the auxiliary attribute values to describe entries, including \emph{column headings} and \emph{row headings} based on their position in the table, as shown in Fig.~\ref{fig:hierarchical-table-composition}.
A \emph{cell} is the intersection of a row and a column, consisting of a single entry.
A \emph{block} is a rectangular group of cells that consists of multiple entries.

\add{The \emph{presentation} of a hierarchical table refers to how cells and blocks are determined using position coordinates.}
Tabular data can be modelled as a two-dimensional matrix.
Users can use the coordinate $\langle$x, y$\rangle$ to denote the cell corresponding to column $x$ and row $y$. 
While $x$ is usually indicated by a capital letter corresponding to the alphabetical order, $y$ is represented by numbers.
For example, the cell $\langle$2, 2$\rangle$ can also be denoted as cell \textit{B2}. 
\add{As for a block, it can be represented by the cell in its upper left corner and the cell in its lower right corner together. For example, the block with upper-left coordinate $\langle$4, 7$\rangle$ and upper-right coordinate $\langle$6, 8$\rangle$ can be denoted as block D7:F8.}

\add{The \emph{logical structure} of a hierarchical table is about associations among labels and entries.
More specifically, the logical structure allows users to use labels rather than positional coordinates to locate the entries.
However, unlike single flat tables, locating an entry along a row/column requires a group of labels since the headings of hierarchical tables consist of multiple levels.
Therefore, it is necessary to determine the relationships between labels.}


Our abstract model of hierarchical tables should accurately specify both logical structures and presentations (\textbf{R1}). In addition, it also needs to facilitate the design of tabular data transformations and visualizations (\textbf{R2}).
\add{Driven by the above requirements, we design an abstract model which focuses on the logical structure of hierarchical tables. The abstract model consists of the following two parts, the relationships between labels and the associations between labels and entries.}

\textbf{The relationships between labels.}
The headings have an inherent hierarchical structure, with several independent or related hierarchies.
When the labels of child nodes in the hierarchies are identical, the abstract model identifies the bidirectional connections between the nodes at different levels and merges these hierarchies as a \textit{bicluster}. 
Compared to independent hierarchies, a bicluster enables users to swap two adjacent levels and update the entry positions in the tabular data correspondingly.
\add{The followings show the formal specifications of hierarchies (denoted as $H$) and biclusters (denoted as $B$) in the abstract model.}

\begin{equation}
\begin{aligned}
B_i = 
\begin{bmatrix}
label_{i,0}\\
\vdots \\
label_{i,n_i}
\end{bmatrix}
\times
B_{i+1},\
B_{l} = 
\begin{bmatrix}
label_{l,0}\\
\vdots \\
label_{l,n_l}
\end{bmatrix}
\end{aligned}
\end{equation}

\begin{equation}
\begin{aligned}
H_{i,j} =
[label_{i,j}]
\times
\begin{bmatrix}
H_{i+1,0}\\
\vdots \\
H_{i+1,n_{i,j}}\
\end{bmatrix},\
H_{l,j}=
\begin{bmatrix}
label_{l,0}\\
\vdots \\
label_{l,n_{l,j}}\\
\end{bmatrix}
\end{aligned}
\end{equation}

\add{For the specification of biclusters, $B_i$ indicates the bicluster start from level $i$, $label_{i,j}$ denotes the inner $j$th label, and $n_i$ refers to the number of labels at level $i$ within bicluster. 
For the specification of hierarchies, $H_{i,j}$ is the $j$th hierarchy starts from level $i$, $label_{i,j}$ is the label of the root node in the hierarchy $H_{i,j}$, $n_{i,j}$ denotes the number of labels in the $j$th subtree, and
$l$ is the number of levels in the headings of tabular data.}
\add{By analyzing if the labels in different hierarchies have same names, HiTailor can automatically identify whether the structure is a bicluster or a hierarchy.}
For example, the row headings ($RH$) in  Fig.~\ref{fig:hierarchical-table-composition} consist of two independent hierarchies and the column headings ($CH$) are modelled as a bicluster, as shown below.
In particular, we use symbols to indicate the statistics by aggregating multiple values. For example, \& refers to the sum of all values at the lower level.

\begin{equation}
{RH}= 
\begin{bmatrix}
[Asia] \times
    \begin{bmatrix}
    [CHN] \times 
    \begin{bmatrix}
    PEK\\
    SHA
    \end{bmatrix},
    [JPN] \times
    \begin{bmatrix}
    OSA\\
    TKY
    \end{bmatrix}
    \end{bmatrix}\\
\\
[{EUR}] \times
    \begin{bmatrix}
    [FRA] \times 
    \begin{bmatrix}
    PAR\\
    MRS
    \end{bmatrix},
    [GBR] \times
    \begin{bmatrix}
    LON\\
    LIV
    \end{bmatrix}
    \end{bmatrix}\\
\end{bmatrix}
\end{equation}

\begin{equation}
CH = 
\begin{bmatrix}
\begin{bmatrix}
2020 \\ 2021
\end{bmatrix}
\times
\begin{bmatrix}
\& \\ spr. \\ aut.
\end{bmatrix}
\end{bmatrix}
\end{equation}




In the formal specifications, labels in headings are arranged from top to bottom (row heading) or left to right (column heading) in the presentation of the hierarchical table.
Therefore, the specification can determine not only the logical structures but also the relative positions between labels in both column and row headings (\textbf{R1}).

\textbf{The associations between labels and entries.}
\add{The abstract model allows users to locate cells or blocks using labels at the bottom level of row and column headings.
For example, the cell ($\langle$131$\rangle$) highlighted in Fig.~\ref{fig:hierarchical-table-composition} can be located with label $SHA$ in the row headings and label $spr$ in the column headings.
However, given that the labels may have the same names, we identify each label at the bottom level using a label sequence from the root node to the target node, denoted by $seq(label)$. For example, label $SHA$ in Fig.~\ref{fig:hierarchical-table-composition} can be identified by $seq(SHA)$ = (\textit{Asia}, \textit{CHN}, \textit{SHA}). The formal specification of $seq(label)$ is defined as follows.}
\begin{equation}
    \begin{aligned}
        seq(label_k) &= (root, \ldots , label_k.parent, label_k)
    \end{aligned}
\end{equation}

\add{Based on this, we define $row~locator$ and $column~locator$ to represent cells and blocks, which are formally defined as follows.}
\begin{equation}
\begin{aligned}
    row/column~locator &= [seq(label_1), \ldots , seq(label_i)]\\
\end{aligned}
\end{equation}

\add{In the above definition, $i$ represents the number of labels used as locators in row/column headings. In particular, locating a single cell requires only one label from row and column headings, respectively. Taking Fig.~\ref{fig:hierarchical-table-composition} as an example, the highlighted cell can be specified as follows:}

\begin{equation}
\label{equ:cell-locator}
\begin{cases}
    ~row~locator = [(Asia, CHN, SHA)] \\
    ~column~locator = [(2020, spr.)]
\end{cases}
\end{equation}

\add{Determining a block often needs more than one label. Note that a block might contain cells from either one hierarchy or across multiple hierarchies. 
In particular, when a block contains all the child nodes, the sequences can be merged with the help of a wildcard (denoted as *).  
Taking Fig.~\ref{fig:hierarchical-table-composition} as an example, $seq(PAR)$ and $seq(MRS)$ can be merged as (\textit{Europe}, \textit{FRA}, *) and Eqn.~\ref{equ:block-locator} below indicates the highlighted block. }

\begin{equation}
\label{equ:block-locator}
\begin{cases}
    ~row~locator = [(Europe, FRA, *)] \\
    ~column~locator = [(2021, *)]
\end{cases}
\end{equation}

The abstract model brings two benefits by using labels to locate entries instead of coordinates.
First, users can change the relative positions of entries by manipulating the labels directly, supporting the interaction design of hierarchical table transformations. 
Second, the relationships between entries are determined by labels in the headings, which serves as the basis for recommendation (\textbf{R2}).

\subsection{Hierarchical Table Transformation}
\label{sec:transformation}

We investigate transformations of hierarchical tables through user interviews and analysis reports for the collected tabular data. 
First, we recruit 15 participants to collect their possible transformations for different hierarchical tables and then use each table in the dataset to conceptualize possible analysis tasks and corresponding transformations. 
We also refer to the analysis reports of statistical experts on data websites and summarize the corresponding transformations for their target analysis tasks.
After that, we examine the applicability of these transformations to other data and remove those operations with low generalizability. 
\add{While the selected operations cannot cover all transformations over hierarchical tables, we carefully choose the common ones that can facilitate the creation of tabular visualizations.}


\add{Embedding visualizations into tabular data has two characteristics. 
First, the underlying data of visualization results is within a contiguous block in the tabular data. 
Second, the size and aspect ratio of the visualization results depend on the area occupied by the corresponding block in tabular data. 
We finally choose six general operations that can facilitate the creation of tabular visualizations and divide these operations into the following three groups.}

\begin{figure}[tb]
 \centering
 \includegraphics[width=0.88\columnwidth]{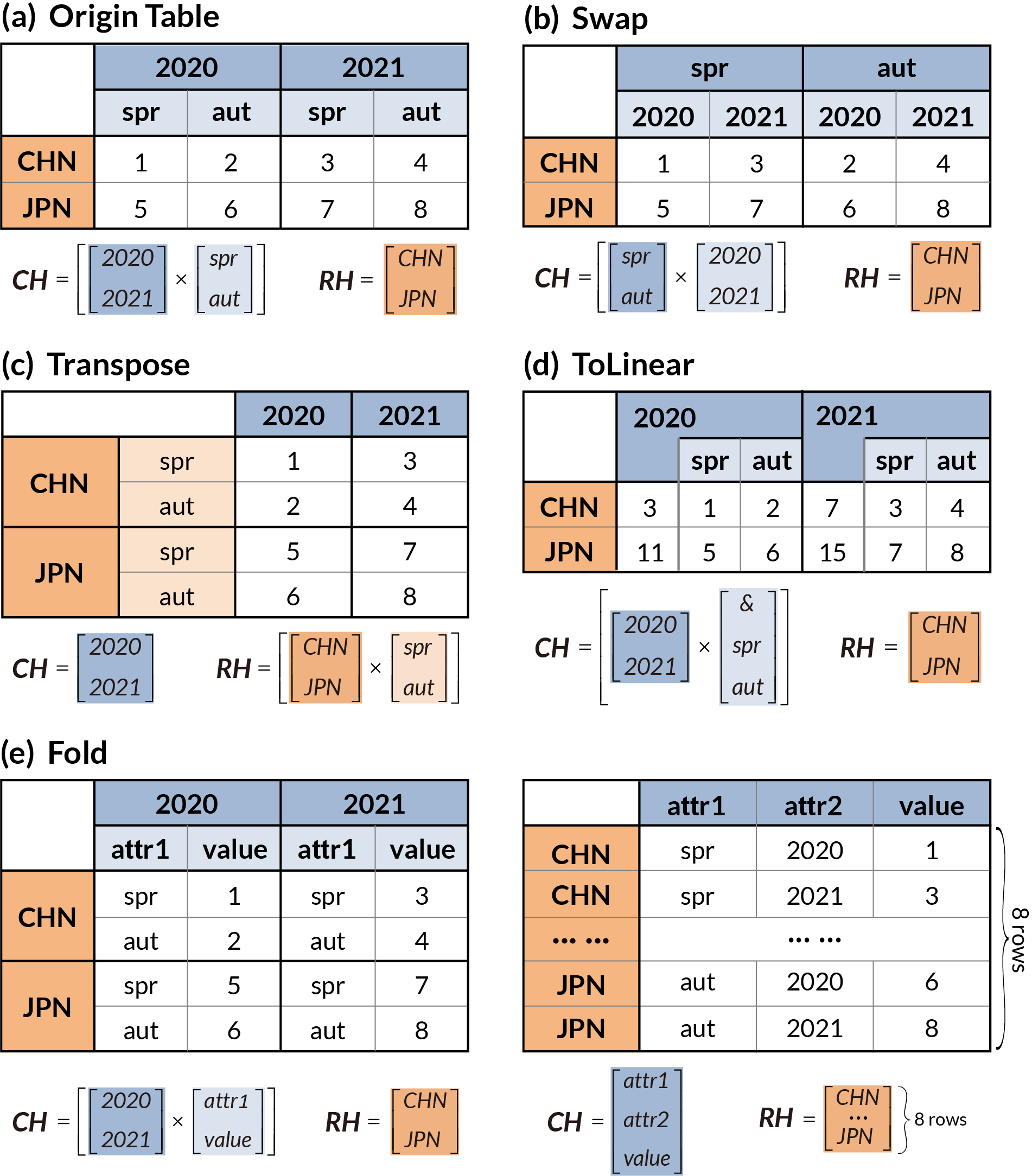}
 \caption{Four different transformations based on the original hierarchical table (a). In particular, the figure above does not illustrate \emph{ToStacked} and \emph{Unfold} operation because these two operations are opposite to \emph{ToLinear} and \emph{Fold} respectively. (e) presents the results after folding the second row and the first row. }
 \label{fig:transformations}
\end{figure}

\textbf{Swap and Transpose}
change the relative positions of labels in column/row headings at different levels and update the inner entries accordingly.
\add{The \emph{Swap} operation is designed to put related entries adjacent to each other within a continuous block, and the \emph{Transpose} operation is for swapping the height and width of blocks.}
The Swap operation enables analysts to change the order of two adjacent levels inside row or column headings, while the Transpose operation supports to exchange row and column headings.
Users can select the appropriate level for transformation according to different analysis requirements to change the position of relevant entries, making comparable entries spatially adjacent to generate visualizations. 
For example, as shown in Fig.~\ref{fig:transformations}, to analyze the changes in data from different years in the same season, users \emph{swap} the two levels in the column headings, as shown in Fig.~\ref{fig:transformations}(b). 
To analyze the variation of data from different seasons in the same country, users \emph{transpose} the second level of the column headings to the row headings, as shown in Fig.~\ref{fig:transformations}(c).

\textbf{ToLinear and ToStacked}
add or delete new nodes to the hierarchy/bicluster and update the corresponding entries without changing the relative positions of labels in the headings. 
\add{Based on the logical structure of headings, the \emph{ToLinear} operation is designed to calculate some statistics for corresponding entries of all nodes in the hierarchy or bicluster, including sum and average, which serve as the underlying data of summary visualizations.
These statistics facilitate users in understanding and analyzing data values from an overview perspective, as shown in Fig.~\ref{fig:transformations}(d).}
However, these derived attributes are usually not comparable to the original values and hinder users' efficient analysis by separating related entries into several discontinuous blocks.
For example, the sum is often much larger than the original values, making the visualization less effective if encoding the sum statistics and other ordinary item values together. 
The \emph{ToStacked} operation allows users to discard these derived attributes from hierarchical tables. 

\textbf{Fold and Unfold}
support transformations between labels and entries.
\add{These two operations are designed to help users rearrange data items in the tabular data to create visualizations that better fit certain aspect ratios.
The \emph{Fold} operation flattens hierarchical tables by converting one row under a multi-level heading into multiple rows under a single-level heading, as shown in Fig.~\ref{fig:transformations}\add{(e)}. 
Conversely, the \emph{Unfold} operation creates new column headings from data values, which can further distinguish between different attributes of a single column.}
\remove{The \emph{Fold} operation flattens hierarchical tables by converting one row into multiple rows, folding a set of columns together into one column and replicating the rest, as shown in Fig.~\ref{fig:transformations}\add{(e)}.
The folding results of hierarchical tables allow users to observe all attribute values of one data item easily.
Conversely, the \emph{Unfold} operation unflattens hierarchical tables, taking two data columns as input and identifying whether each column is categorical or numerical based on its characteristics.
Furthermore, it extracts rows with the same values in all the other columns and unfolds the two chosen columns.}
\add{In addition, it transforms the repeated entries into labels in the headings so that the original data, which is arranged only horizontally or vertically, can be arranged on the whole two-dimensional space based on the logical structure.
Because the area with a high aspect ratio is not suitable to create various visualizations (\textit{e.g.}, scatter plot), \emph{Unfold} operation can rearrange the original data in a rectangular area with a better aspect ratio.
For example, Fig.~\ref{fig:transformations}(a) shows the result after applying the unfold operation to the right table of Fig.~\ref{fig:transformations}(e). More specifically, all data entries in the table of Fig.~\ref{fig:transformations}(a) are originally arranged within the single column (\textit{value}) in the table of Fig.~\ref{fig:transformations}(e). }






\subsection{TableUnit Recommendation}
\label{sec:recommendation}

We define a cell/block specified by users in the hierarchical tables as a \emph{TableUnit}, which is regarded as the primary component for tabular visualizations. To prevent users from repeatedly selecting multiple TableUnits and specifying the same visualization configurations, it is necessary to recommend TableUnits that are logically related to the user-selected one.
In this section, we introduce different mechanisms for TableUnit recommendation.

The recommendation for TableUnits is based on the logical structure of hierarchical tables.
According to Sec.~\ref{sec:abstract-model}, the logical structure can be represented as relationships between labels. Therefore, our method describes a TableUnit using labels, in order to facilitate the recommendation process.
\add{More specifically, we define \emph{row descriptor} and \emph{column descriptor} for each TableUnit, using the information from locators of the cell/block corresponding to the TableUnit.
A descriptor is defined as a set of labels, which consists of the last non-wildcard label of all sequences from the corresponding locator.}
Taking Fig.~\ref{fig:hierarchical-table-composition} as an example, the row and column descriptors of the block located with Eqn.~\ref{equ:block-locator} are [\emph{FRA}] and [\emph{2021}], respectively.

\begin{figure}[tb]
\centering 
    \includegraphics[width=0.85\columnwidth]{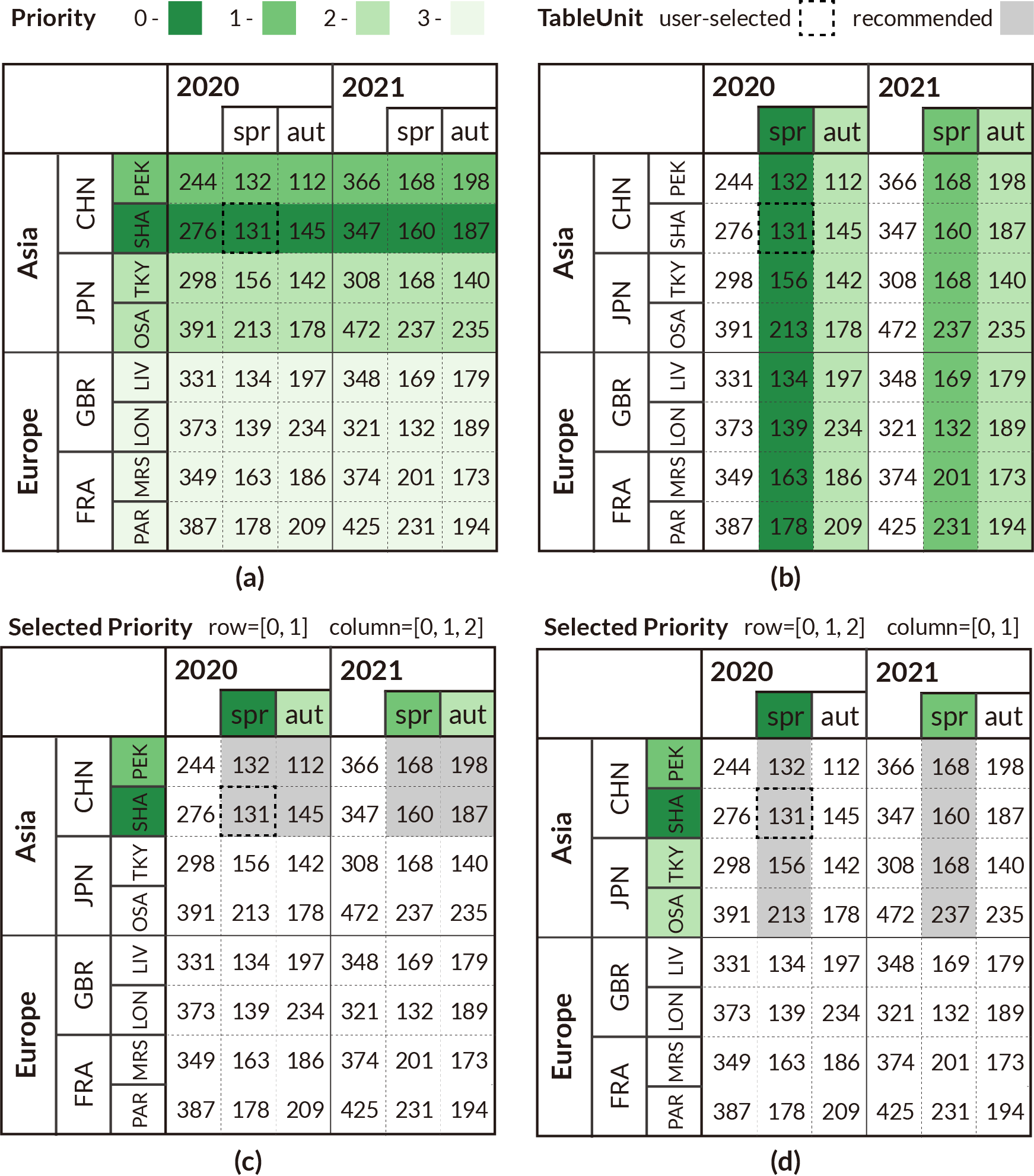}
    \add{\caption{The two figures in the first row show two different methods to calculate the priority of descriptors. (a) shows the topology-based mechanism and (b) shows the name-based mechanism. The two figures in the second row show different TableUnits for recommendation according to the priority range specified by users.}}
    \label{fig:priority}
\end{figure}

\add{After users select a TableUnit, our method computes the \emph{priorities} of other related TableUnits to describe their relationships and then generates recommendations automatically.}
To calculate the priority of TableUnits, we first design the priorities of its descriptors. 
Since a TableUnit has both row and column descriptors, we define their priorities respectively, denoted as \emph{row priority} and \emph{column priority}.

The priority of a descriptor indicates its correlation with the descriptor of the user-selected TableUnit, where a higher priority indicates a stronger relevance.
\add{Based on the abstract model explained in Sec.~\ref{sec:abstract-model}, the correlations between labels are reflected in two aspects. First, hierarchies establish the topological relationships between sibling labels. Second, biclusters complement the relationships between labels with the same name. 
Therefore, we generate two types of priority computation methods for descriptors (\emph{i.e.}, \emph{topology-based} and \emph{name-based}).}

In both of these two methods, descriptors of the user-selected TableUnit are identified as \emph{reference descriptors}.
Taking the user-selected TableUnit shown in Fig.~\ref{fig:priority} as an example, the reference descriptors include [\textit{SHA}] from row headings and [\textit{spr}] from column headings.
We assign specific numbers to descriptors as their priorities, where a smaller number indicates a higher priority.
Therefore, the priority of a reference descriptor is assigned as 0.
It is worth noting that row and column priorities are calculated separately.
The details of these two mechanisms are described as follows.

\textbf{Topology-based Mechanism.}
\add{When computing the priority of a descriptor $d$, the topology-based method finds the lowest common ancestor (denoted as \textit{lca}) of $d$  and its reference descriptor (denoted as \textit{ref}) based on the abstract model. 
Then we use the level of \textit{ref} and \textit{lca} (denoted as $L_{ref}$ and $L_{lca}$) to calculate the priority number of $d$ (denoted as $P_d$), as shown in Eqn.~\ref{equ:topo-priority}. In particular, we assign $L_{lca}$ as 0 if the \textit{lca} does not exist.}
\add{For example, the reference descriptor in Fig.~\ref{fig:priority}(a) is [\textit{SHA}], and the lowest common ancestor of [\textit{TKY}] (at level 3) and [\textit{SHA}] is [\textit{Asia}] (at level 1). Therefore, the priority of [\textit{TKY}] is $2$.} 
\begin{equation}
\label{equ:topo-priority}
    \begin{aligned}
        P_d ~= ~L_{ref} ~- ~L_{lca}
    \end{aligned}
\end{equation}

\textbf{Name-based Mechanism.}
\add{When calculating the priority of a descriptor $d$, the name-based method focuses on the label of both $d$ and the reference descriptor (denoted as \textit{ref}). 
If $d$ and \textit{ref} have exactly the identical label, the priority of $d$ is assigned as 1. Otherwise, the priority is assigned as 2.}
\add{For example, as shown in Fig.~\ref{fig:priority}(b), the reference descriptor from column headings is [\textit{spr}]. As a result, the priority of other descriptors with the name \textit{spr} is assigned as 1, while the priority of other descriptors is assigned as 2.}

Based on the priority computation methods of TableUnit descriptors mentioned above, our method allows users to identify the priority of a TableUnit through its row/column priority.
After selecting a TableUnit, users can choose a method to compute the row and column priority, including the topology-based method and the name-based method.
\add{Furthermore, they can select the priority range of both row and column, and then HiTailor will recommend TableUnits according to the selected range.
For example, Fig.~\ref{fig:priority}(c) and (d) shows the recommended TableUnits according to users' specified priority range. Note that HiTailor does not make recommendations when the selected TableUnit does not conform to be within a single subtree from both row and column headings, because selecting such TableUnit is usually used to analyze specific data items.}

\subsection{TableUnit Visualization}

After selecting TableUnits, users need to visualize the selected data items and embed the visualization results into hierarchical tables. 
Because TableUnit is the primary component for tabular visualizations, we introduce the visualization techniques for a single TableUnit in this section. 
In addition, we also discuss how to apply the visualization specifications of the selected TableUnit to other recommended ones.

\subsubsection{TableUnit Decomposition}
\label{sec:unit-decomposition}

\begin{figure}[tb]
 \centering
 \includegraphics[width=0.9\columnwidth]{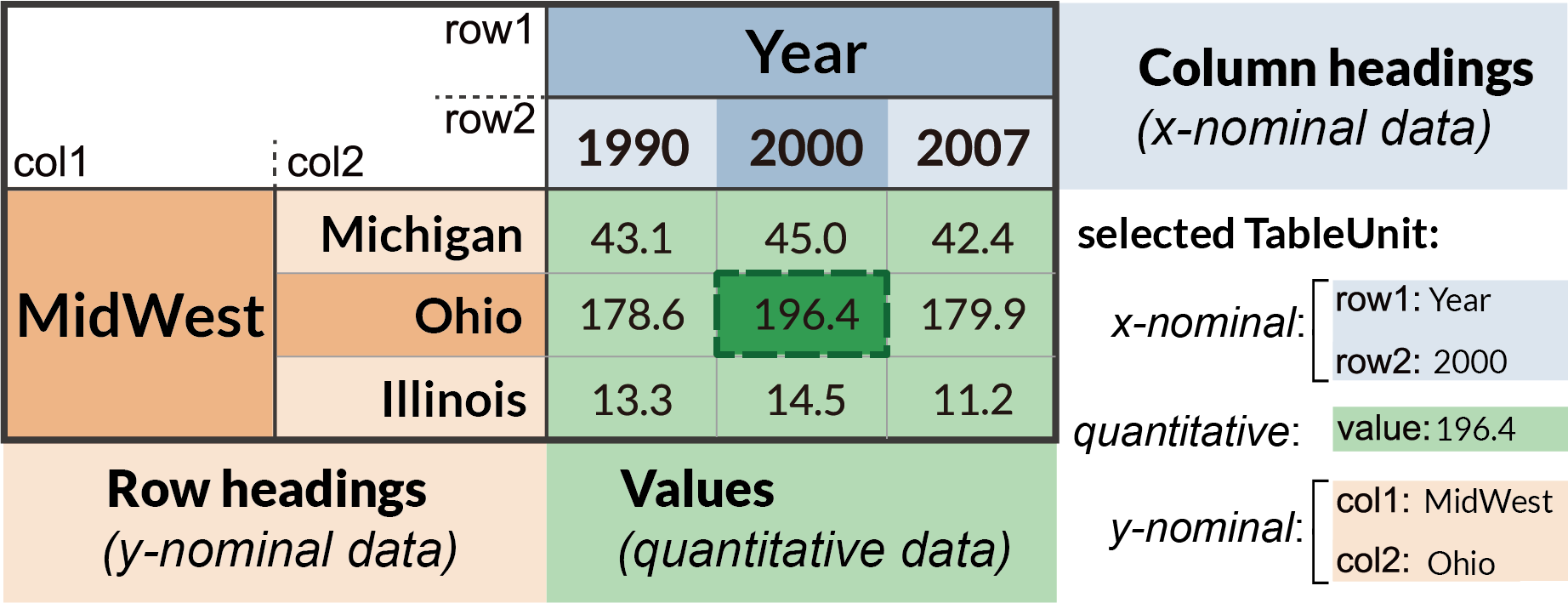}
 \caption{The decomposition of a TableUnit in hierarchical table consists of three components, \emph{x-nominal} data, \emph{y-nominal} data, and quantitative values.}
 \label{fig:tbm}
\end{figure}

A TableUnit selected by users is either a cell or a block.
According to the abstract model explained in Sec.~\ref{sec:abstract-model}, a cell or a block is specified by its corresponding labels. For example, each entry in Fig.~\ref{fig:tbm} has five attributes. \add{One is the data value of entry, and the others come from the labels used to locate the entry, including row headings and column headings.}
Typically, the labels are in string format (e.g., name or date), and the data value is a quantitative number. We define the attributes from labels as nominal data and the data value as quantitative data. 

Tabular visualization techniques embed the visual representations into hierarchical tables. Therefore, associating visual elements with their underlying data items could facilitate users' understanding and exploration. 
When a TableUnit contains only one cell, TVs retain a tabular layout and place the visual element within the cell.
However, when a TableUnit contains multiple cells, the visual elements' positions in the visualization results might encode its attribute values. 
According to the above decomposition method, a TableUnit with multiple cells will have a sequence of attributes from column/row headings as nominal data. 
To make the visual elements in the visualization results align with the headings, the decomposition results of TableUnit keep the attributes sequence. 
We define the attributes from row headings as vertically arranged in a strict order called \emph{y-nominal} data, which is similar to row headings. 

\subsubsection{Visual Mapping Mechanism}
To improve the intuitiveness of visualization results, we provide restricted configurations about visual mapping. 
As mentioned above, we decompose the attributes of TableUnits into three categories, \emph{x-nominal} along with the horizontal direction, \emph{y-nominal} along with the vertical direction, and the \emph{quantitative} values within the cell. 
Most visualization techniques have one or multiple axes arranged in horizontal/vertical direction. 
Creating TableUnit visualization requires users to encode different attributes of data items to horizontal/vertical positions along the x/y axis in the configuration. 
The visual mapping does not enable users to bind the attributes of TableUnit to any axis arbitrarily. More specifically, it only allows users to bind \emph{x-nominal} data to the horizontal axis or \emph{y-nominal} data to the vertical axis.

In addition, we carefully design feasible parameters of visual mapping mechanisms, which realize the reusability of visualization specifications.
For hierarchical table visualization, users need to reuse the configuration specification of the selected TableUnit for others with a similar logical structure. 
However, the underlying data change completely for different TableUnits. Therefore, the attributes in the original specification should change accordingly. 
Our visual mapping mechanisms decouple visualization configuration with the underlying data.
For instance, a user creates a stacked bar chart by encoding the \emph{x-nominal} data to horizontal positions, the value to height, and \emph{y-nominal} data to color.
Then, the user wants to apply this configuration to other TableUnits. 
Our method analyzes the underlying data, and binds the corresponding \emph{x-nominal} data, data value, and \emph{y-nominal} data to the same visual channels, getting a logically coherent visualization result.

\subsubsection{Visualization Templates}

We choose several visualization techniques for TableUnits in hierarchical tables and divide these visualizations into the following four categories to meet users' requirements of tabular data analysis.
In particular, unit visualization is designed for the TableUnit with only one cell. The other three types of visualizations belong to summary visualization, and are suitable for TableUnits with multiple cells.

\textbf{Unit Visualization} visualizes the value of each cell, showing the relative magnitude between comparable cells without changing the structure of a table.
It normalizes values for one user-selected cell, and other related cells determined by the recommendation mechanism explained in Sec.~\ref{sec:recommendation}. 
Users can encode the value of each cell with different visual channels such as color, size and position. The corresponding visual elements will be embedded in each cell when users determine the specifications. 
The unit visualization creates a handy way to perceive the relative value of tabular data. For instance, by encoding value with the color density, users can learn the cells with the maximum or minimum value according to the color shading like a heatmap. 

\textbf{Data Overview} is designed to show the values of all cells within TableUnit. 
The visualization techniques for data overview consist of bar chart, stacked bar chart, ranged dot plot, box plot, strip plot, parallel coordinate, multi-series line chart, pie chart, and radial plot. 
We divide the data overview technique into two categories according to whether aggregating values in TableUnit.
The techniques without aggregation encode each value in TableUnit into a visual element and compute the position based on the cell's value.
For instance, the stacked bar chart arranged along the horizontal direction binds the \emph{x-nominal} data to the x-axis, the \emph{value} to the height, and the \emph{y-nominal} data to the color channel. 
However, aggregation techniques compute a single value (such as minimum, maximum, average) from multiple values in a row/column and then encode it to the visual element.
For example, the ranged dot plot along the horizontal direction binds the \emph{x-nominal} data to the x-axis and encodes the minimum and maximum values from the column to the position of visual elements. 

\textbf{Trend Tracking} enables users to understand variance along the column or row. It encodes the sequence of headings and the values of one row or column to support users learning the trend. The visualizations for trend tracking consist of the horizon graph and line chart. The horizon graph is appropriate for showing the trend tracking visualization because it is designed to encode the value in a space with limited height like the table cells~\cite{2009-horizon}. Users can bind the column headings (e.g., \emph{years}) to the \emph{x-axis} and bind values of one row (e.g., \emph{France}) to the \emph{y-axis}, and then they can learn the changes in financial transactions according to the variance of height and color.

\textbf{Correlation Exploration} allows users to explore the correlations between multiple rows or columns within the TableUnit. The visualizations about correlation exploration consist of scatterplot and heatmap, which requires users to bind the quantitative values to the axis. More specifically, users can specify the data organization structure to column or row at first and then encode the data to the different axis to validate whether some rows or columns have correlations.

\section{The HiTailor System}
We have designed and implemented a prototype system to support users authoring the tabular visualization for hierarchical tables interactively. 

\subsection{Design Principle}
The prototype aims to enable users to create the visualization results for hierarchical tables interactively. We identified the following four design principles for the prototype system.
\add{These four principles are designed to reduce users' burdens of creating tabular visualizations and analyzing hierarchical tables in different scenarios using HiTailor.}

\textbf{DP1: Support both presentations and explorations for tabular data.}
Data presentation means creating static visualizations to communicate insights, while exploration indicates creating visualization to understand data and find insights interactively~\cite{2019-taggle-iv}.
To support the explorations, HiTailor enables interactive transformations and visualizations. 
To support the presentations, HiTailor is able to export the visualization results for tabular data as static images for further editing. 



\textbf{DP2: Build the correlations between the data items and visual elements in tabular visualization.}
The benefit of the tabular visualization technique is reducing the cognitive burdens by integrating data with visualizations, because users do not need to switch their attention between visualizations and data values~\cite{2006-AyersSweller}.
More specifically, it replaces the data values in tabular data with visualizations, which help users understand their correlations.
For unit visualization, each cell serves as the basic TableUnit and users can easily understand detailed relationships. 
However, it is difficult for users to identify the complex relationships in summary visualizations generated based on data blocks consisting of multiple cells. 
The prototype system supports the interactive highlighting between the visual elements and the underlying row/column headings. 
In addition, hovering on the visual elements can check the detailed data attributes of the underlying data items. 



\textbf{DP3: Balance the direct manipulations and configuration panels.}
Direct manipulation is a natural and intuitive interaction approach. However, it is difficult for users to specify complex parameters precisely. 
The goal of HiTailor is to enable analysts to author tabular visualizations with minimal difficulty and tedium.
Therefore, HiTailor provides both direct manipulation and configuration widgets.
HiTailor supports the direct manipulation interaction for tabular data transformations. 
Direct manipulations have a large operational space, and users might be overwhelmed. HiTailor provides several visual cues as the guidelines to suggest possible transformations for users. 
In addition, HiTailor provides configuration panels for several parameter widgets based on Vega-Lite to allow users to adjust the values directly. 
Furthermore, to facilitate users associating the parameters and row/column headings, the specific row or column will be highlighted when selecting the detailed parameters in the widgets.

\textbf{DP4: Reduce the cognitive burden for constructing visualizations for the TableUnit.}
As mentioned above, block visualizations of TableUnit are supported by Vega-Lite, a high-level grammar of interactive graphics. 
It provides a declarative JSON syntax to create an expressive range of visualizations for data analysis and presentation.
Compared to the imperative programming methods, Vega-Lite reduces users' cognitive burden of visualization construction. However, it still consists of many parameters to support its expressiveness. 
To improve the efficiency of visualization construction, HiTailor utilizes the template editor technique authoring approach. Specifically, it contains the initial step of visualization template selection and the refinement of the selected visualization. HiTailor makes simplifications for the parameters configurations panel in the refinement step.
\subsection{User Interface and Interaction}
The user interface of HiTailor consists of the transformation operator panel, tabular visualization panel, visualization template panel, and visualization configuration panel, as shown in Fig.~\ref{fig:teaser}. These four interactive panels support the following process of tabular data visualization.

First, users upload a hierarchical table, and then the HiTailor prototype system presents it in the tabular visualization panel and parses the tabular data into an abstract model. 
HiTailor identifies hierarchical/biclustering structures from the row and column headings according to the relative position of cells. 
Then users can make transformations for the hierarchical table through interactive manipulations (e.g., \emph{drag} and \emph{drop}) or click the operators in the transformation operator panel directly (\textbf{DP3}).
In particular, users' transformations for hierarchical tables also change the underlying abstract model accordingly.

Based on the transformation results of hierarchical tables, users can specify a cell/block within the tabular data as a TableUnit. 
According to users' selections and the abstract model, HiTailor recommends other related TableUnits with different priorities, shown in the tabular visualization panel. 
More specifically, the priority of different TableUnits is encoded into the color density, and users can adjust the selected range of TableUnit to apply the same visualization template. 

For TableUnit visualization, users begin by choosing a visualization technique from the visualization template panel and update the specifications and detailed parameters in the visualization configuration panel.
Taking users' selected data blocks as input, the configuration editor shows the preview of visualization results based on Vega-Lite (\textbf{DP4}). 
When users are satisfied with the preview of visualization results, they can apply the visualization template to all TableUnits (user-selected and recommended) in the tabular visualization panel.

Note that tabular visualization authoring is an iterative process. Users can specify other blocks as TableUnit and apply different visualization templates to these blocks. 
Based on the visualization results of hierarchical tables, users can explore the data interactively, including checking the underlying data of the visualization and highlight other related data items in the table (\textbf{DP2}). 
When users are satisfied with the tabular visualization results, they can export the results in different formats to support flexible presentations of hierarchical tables (\textbf{DP1}).

\section{Use Case: Game Hardware Sales Analysis}
We validate the effectiveness of HiTailor through a case study. The analysis is carried out by analysts of the game market, using the sales data of game hardware. The results of the case study show that HiTailor facilitates users to gain insights from hierarchical tables, which is a common task in business intelligence.

\add{The use case is based on the game hardware sales dataset from VGChartz\footnote{www.vgchartz.com}, a website that weekly estimates hardware and software sales in the video gaming industry.
The analysts collect the historical sale data items, which record the sales of consoles from different manufacturers in several regions for each quarter from 2013 to 2020.
Then they construct a hierarchical table in the \textit{xlsx} format, widely used for Excel. The column headings of the hierarchical table are about quarters and years, and the row headings are about consoles from different manufacturers in several regions.}
Multiple factors influence the sales of game hardware, and one of the essential aspects is software. 
However, the preferences of consumers in different regions vary a lot. 
Therefore, the release of game software may impact hardware sales differently.
In addition, the changes in hardware sales are also influenced by the prices, the release of new products, etc.
To explore the patterns for hardware sale variance and find explanations, the analysts should comprehensively analyze different aspects of hierarchical tables.




The analysts load the tabular data of hardware sales, consisting of 32 columns and 42 rows, into the HiTailor prototype system. 
To explore the relationships between time and sales, the analysts apply the \emph{Swap} operation for two rows about years and seasons to reshape the hierarchical table and the \emph{ToLinear} operation to get the overall sales of every platform. 
\add{HiTailor allows users to select different parts of tabular data to apply different visualizations to build an overview, with keeping the table structure. The analysts first select the first row (Nintendo, Nintendo 3DS, \&) of the hierarchical table and use the horizon graph as shown in Fig.~\ref{fig:teaser}(b3), a visualization for trend tracking, to show the variance of overall sales over time. }
After applying this visualization template to all recommended blocks, the analysts can summarize and compare the trends of different hardware.
The analysts get two findings from the visualization results: First, unlike other electronics, the sales of game hardware do not hit the highest point in the first year. The sales peaks in the middle of the hardware’s life cycle and then decreases gradually.
Second, the sales of game hardware fluctuate periodically. 
All hardware sales increase at the end of each year dramatically because of promotion events on Black Friday. 
In addition, the PlayStation Vita (PSV) of Sony also has a sales spike around September 2015, mainly because of the release of popular game software on this hardware. 

Next, the analysts want to analyze the sales of different products in different regions.
They choose the unit visualization to show the first four products of Nintendo. More specifically, it maps the selected cells to rectangles and encodes the value to size and color.
\add{From the visualization results (Fig.~\ref{fig:teaser}(b4)), the analysts find that the sales of Nintendo 3DS (3DS), Nintendo DS (DS), and Wii (Wii) are similar in Europe, Japan, and North America. However, the sales of Nintendo Switch (NS) in North America are better than in Europe and Japan. 
This validates that the Nintendo Switch (NS) is more popular than the other three hardware.}
\add{The visualization results explained above focus on the variance of absolute numerical values. 
Furthermore, to explore the sales proportions among different regions of Xbox One and PlayStation Vita, the analysts select the radial plot, a visualization template belonging to the data overview category. 
The visualization results (Fig.~\ref{fig:teaser}(b6)) show that the sales of Xbox One mainly come from North America, while the gaming market in Japan almost has no interest in Microsoft's consoles, and the sales of PlayStation Vita are on the contrary. }

After confirming consumers' distinct preferences among different regions, the analysts use the box plot to explore consumers' purchasing behaviors for PlayStation 3 (PS3) and PlayStation 4 (PS4). 
\add{According to the visualization results shown in Fig.~\ref{fig:teaser}(b5), sales of two hardware in Europe and North America are better than in Japan, especially for PlayStation 4 (PS4).
Another finding is that the box plot about the sales of North America always consists of an outlier, which indicates an exceptionally high sale record.
After checking the values in the raw table, the analysts found that all these outliers indicate the sales in the last quarter. It is because of the promotion events explained above. 
However, no outlier exists in the boxplot of Japan's sales, which inspires that chain retail stores should have different stocking strategies for these regions throughout the year.}
\vspace{-0.1cm}
\section{\add{User Study}}
\label{sec:user-experiment}
\add{We conducted a comparative study to evaluate the effectiveness of HiTailor for analyzing hierarchical tables.
To make it a fair experimental comparison, we chose baselines from existing tools with similar functions to HiTailor, which support table transformation and visualization generation.
Tableau and Microsoft PivotTable are two well-known tools for interactive tabular data analysis. PivotTable supports powerful tabular data transformations but fails to provide in-situ visualization in the table, i.e., directly visualizing the transformed data in the hierarchical tables. However, HiTailor adopts similar workflows with HiTailor that integrate tabular data transformation and visualization seamlessly.
Therefore, we chose Tableau as a reasonable counterpart with HiTailor.}

\vspace{-0.1cm}
\subsection{\add{Experiment Dataset}} 

\add{In the experiment, we used the artificially generated tabular data for better control of the experiment. To help users understand the data and the analysis tasks, we generated an experiment dataset positioned in the semantic context of industrial productivity over time. Specifically, the tabular data are designed with the row headings `cities' and `industries' and the column headings `years' and `quarters'. Multiple datasets are sampled from this tabular frame, which have the same size and tabular structure but different data values. } 



\begin{figure}[tb]
 \centering
 \includegraphics[width=0.9\columnwidth]{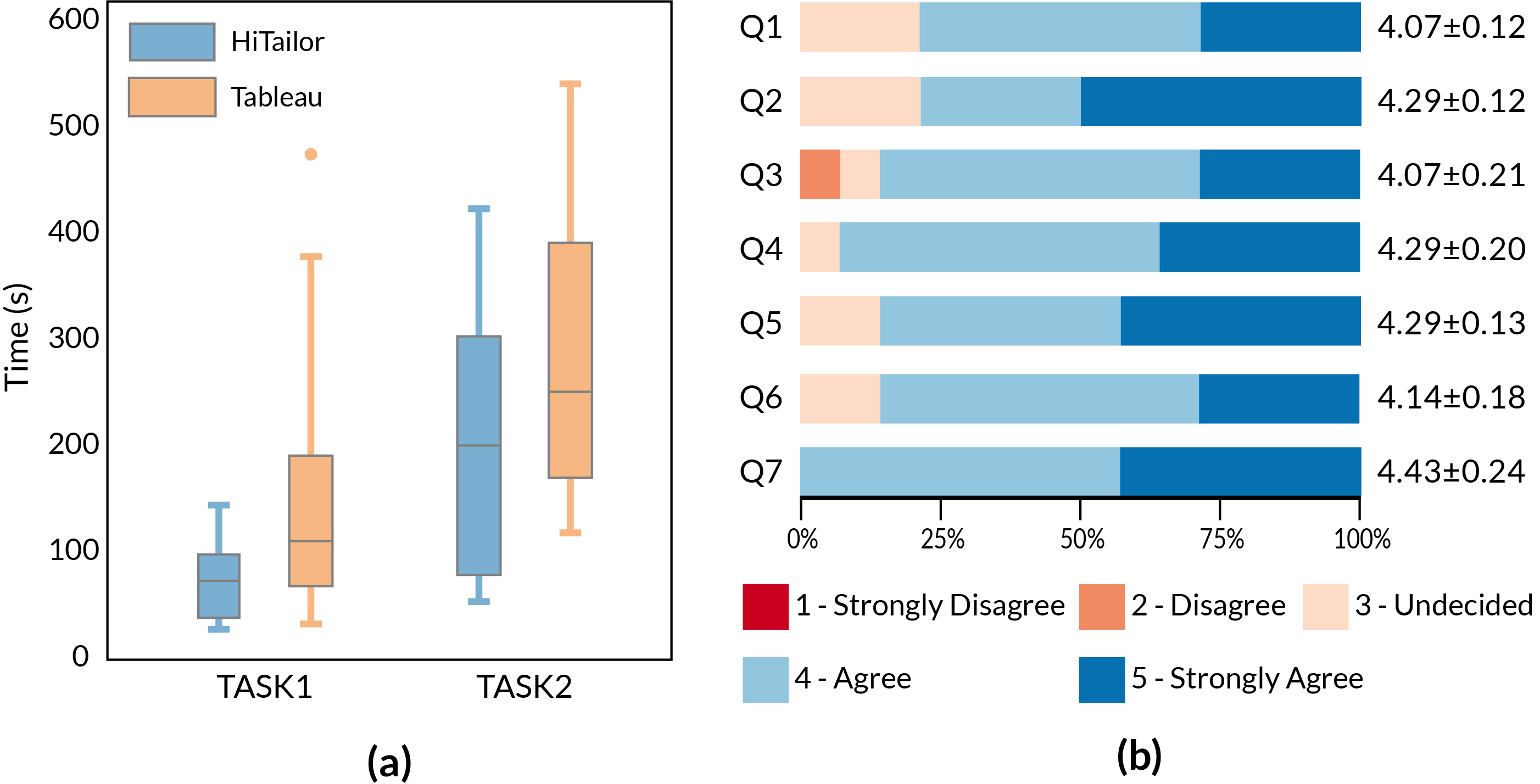}
 \caption{The result of the user study. (a) shows the time distribution of participants using HiTailor and Tableau to complete different tasks. (b) shows the ratings from different aspects of HiTailor on a five-point Likert scale (N = 16). The rightmost column indicates the average and standard deviations. (Q1: Easy to Learn. Q2: Easy to use. Q3: Interactions intuitive. Q4: Transformations helpful. Q5: Recommendation helpful. Q6: Visualization satisfactory. Q7: Facilitate data understanding.)}
 \label{fig:user-study-result}
\end{figure}

\vspace{-0.1cm}
\subsection{\add{Experiment Tasks}}

\add{We choose two tasks for the comparative study, mainly referring to the process of visual information seeking~\cite{1996-eyes-ben}, which includes obtaining an overview (Task 1) and exploring data in detail (Task 2). In particular, the headings of the hierarchical tables consist of labels on multiple levels. One characteristic of tabular data is that the inner entries may be associated with headings at different levels. Therefore, the second task is to simultaneously explore the data entries at different levels.}

\add{\textbf{Task 1}: This task requires users to understand the total GDP of each city in different years (the trend of GDP over time). Specifically, users are asked to identify the city whose trend is different from the others.}

\add{\textbf{Task 2}: This task is about analyzing the total and each industry's GDP for different cities. 
Users first need to identify the years when the total GDP reaches the maximum value, and then compare the portion of each industry's GDP over the above years. Specifically, users are asked to indicate the industry with the most significant portion.}

\subsection{\add{Experiment Configuration}}

\add{\textbf{Participants and Apparatus.} 
We recruited 16 participants, including 6 females and 10 males. All of them were undergraduate or postgraduate students with different majors, ranging from computer science to mathematics and statistics, and they often use table to organize and analyze data in their practical applications. 
The user study was performed in a quiet computer lab on a Dell Precision T5500 desktop PC, with an Intel Xeon Quad-Core processor, 8GB RAM, and an NVIDIA Quadro 2000 graphics card driving a 23-inch LCD 1920 $\times$ 1080 pixel monitor.}

\add{\textbf{Procedure.} 
The experiment session lasted around one hour. 
First, we provided a tutorial about HiTailor and Tableau, including the functionalities of different transformations and the interactions for creating tabular visualizations. 
We asked the participants to perform two tasks mentioned above on the generated datasets using Tableau and HiTailor, respectively, and record their results and the time of completion.
Within each task, users underwent each of the two techniques in counterbalanced order.
At the end of the experiment, we asked the participants to rate HiTailor from different aspects using a five-point Likert scale.
Note that we asked users to complete each trial as accurately and quickly as possible and informed them that accuracy and completion time are equally important.
We also conducted a follow-up interview for the participants for about 10 minutes to collect their feedback on HiTailor. }

\subsection{\add{Experiment Results}}
\add{We analyzed the completion time and accuracy for each task. 
Through the analysis of variance, we found that using different techniques had a significant effect on time (Task1: $F=5.82, ~p<0.05$; Task2: $F=9.85, ~p<0.01$). 
According to the result of the t-test, HiTailor is significantly faster than Tableau in both tasks (Task1: $p<0.05$; Task2: $p<0.01$), as shown in Fig.~\ref{fig:user-study-result}(a). 
As for the accuracy, Task1 had an accuracy rate of 78.57\% using Tableau and 92.86\% using HiTailor, while Task2 had an accuracy rate of 64.29\% using Tableau and 85.71\% using HiTailor.
Then, we analyzed the results of user ratings on the usability of HiTailor, as shown in Fig.~\ref{fig:user-study-result}(b). 
We found that HiTailor received a high rating from participants about facilitating the understanding of the data ($\mu=4.43,~\sigma=0.24$). Many participants agreed on the usefulness of our transformations ($\mu=4.29,~\sigma=0.20$) and recommendation ($\mu=4.29,~\sigma=0.13$).}

\vspace{-0.15cm}
\section{Discussion and Future work}

\textbf{Scalability}. 
\add{The abstract model of a hierarchical table can accurately determine its presentations. Therefore, the computational complexity of data transformation is O($n$), and $n$ denotes the number of entries in tabular data.}
In addition, we optimize the rendering mechanism to improve the scalability of HiTailor for large tabular data. More specifically, the system only renders the visible part of the tabular data. 
The implementation of the HiTailor system is based on Scalable Vector Graphics (SVG) and the rendering performance will decrease due to a large number of DOM elements. In the future, we plan to mitigate this problem by replacing the DOM-based rendering with an HTML5 Canvas implementation.  

\add{\textbf{Evaluation}. 
We present the user experiment, which compares the effectiveness of completing tabular data analysis tasks using HiTailor and Tableau in Sec.~\ref{sec:user-experiment}. 
Considering Tableau and HiTailor are both designed for facilitating users' analysis of hierarchical tables, we choose the subjects in these related areas, including computer science, mathematics and statistics. 
These subjects use tables for data processing and management or have experience with data visualizations.
For future work, we will continue to investigate the analysis of tabular data for general users and validate our approach's effectiveness for a wider population range.}

\textbf{Visual encoding consistency}. The authoring of table visualizations with HiTailor is an iterative process, as shown in Fig.~\ref{fig:pipelineX}. 
Therefore, the table visualization consists of different TableUnits specified by users many times, and the visualization specifications of these TableUnits are independent.
From the perspective of the whole tabular visualization, assigning the same visual channel (e.g., color) to semantically unrelated attributes can confuse users. 
However, depending on the number of independent TableUnit, it might not be possible to choose different visual channels for all TableUnits in the tabular visualization results.
Therefore, HiTailor allows users to reuse the visual channels and improve tabular visualization results' readability. Users can view ranges with consistent visual mappings interactively.

\textbf{Intelligent transformation and visualization}. HiTailor enables users to present and explore hierarchical tables through transformations and visualizations. 
The visualization process for the hierarchical tables highly relies on users' interaction. 
Future work could explore more intelligent tabular visualization techniques of hierarchical tables. 
First, it requires defining metrics to evaluate the transformation results.
In addition, we plan to take the semantics of column/row headings of tabular data into consideration by using the knowledge graph techniques. 
To reduce the cognitive burden for authoring visualizations, we also want to explore the interaction approaches based on natural language.


\vspace{-0.1cm}
\section{Conclusion}
We have presented HiTailor, a technique for authoring tabular visualizations of hierarchical tabular data. HiTailor is based on an abstract model for hierarchical tables, which defines its column/row headings as hierarchical and biclustering structures. HiTailor implements three pairs of transformation operators based on the abstract model and enables users to transform the hierarchical tables interactively. 
Users can specify a data cell/block within the body of a hierarchical table as TableUnit, and HiTailor will recommend the other TableUnits according to different mechanisms to improve the efficiency of visualization construction. 
We demonstrate the utility of the HiTailor system through a qualitative study and a comparative study.
The result shows that the HiTailor system can enhance users' analysis and understanding of hierarchical tables.
\add{HiTailor is available at \href{https://github.com/bitvis2021/HiTailor}{https://github.com/bitvis2021/HiTailor}.}


\acknowledgments{
We thank the anonymous reviewers for their valuable comments. This work is supported by the National Key Research and Development Program of China (2021YFB3301500) and Beijing Institute of Technology Research Fund Program for Young Scholars.}

\bibliographystyle{abbrv-doi}

\bibliography{tablevis}

\begin{thebibliography}{10}

\bibitem{2015-fuse-barik}
T.~Barik, K.~Lubick, J.~Smith, J.~Slankas, and E.~Murphy-Hill.
\newblock Fuse: A reproducible, extendable, internet-scale corpus of
  spreadsheets.
\newblock In {\em Proc. IEEE/ACM Conf. Mining Software Repositories (MSR)}, pp.
  486--489, 2015.

\bibitem{2015-FlashRelate-Barowy}
D.~W. Barowy, S.~Gulwani, T.~Hart, and B.~G. Zorn.
\newblock Flashrelate: extracting relational data from semi-structured
  spreadsheets using examples.
\newblock In {\em Proc. ACM SIGPLAN Conf. Programming Language Design and
  Implementation (PLDI)}, pp. 218--228, 2015.

\bibitem{2006-integrated-bautista}
J.~Bautista and G.~Carenini.
\newblock An integrated task-based framework for the design and evaluation of
  visualizations to support preferential choice.
\newblock In {\em Proc. Conf. Advanced Visual Interfaces (AVI)}, pp. 217--224,
  2006.

\bibitem{2016-matrix-behrisch}
M.~Behrisch, B.~Bach, N.~Henry~Riche, T.~Schreck, and J.-D. Fekete.
\newblock Matrix reordering methods for table and network visualization.
\newblock {\em Computer Graphics Forum}, 35(3):693--716, 2016.

\bibitem{1977-interactive-benson}
W.~H. Benson and B.~Kitous.
\newblock Interactive analysis and display of tabular data.
\newblock In {\em Proc. Conf. Computer Graphics and Interactive Techniques
  (SIGGRAPH)}, pp. 48--53, 1977.

\bibitem{1973-semiologie-bickmore}
J.~Bertin.
\newblock {\em S{\'e}miologie Graphique: les diagrammes, les r{\'e}seaux, les
  cartes}.
\newblock {\'E}ditions Gauthier-Villars, Paris, 1973.

\bibitem{2004-ValueCharts-Carenini}
G.~Carenini and J.~Loyd.
\newblock Valuecharts: analyzing linear models expressing preferences and
  evaluations.
\newblock In {\em Proc. Conf. Advanced Visual Interfaces (AVI)}, pp. 150--157,
  2004.

\bibitem{2022-rigel-chen}
R.~Chen, D.~Weng, Y.~Huang, X.~Shu, J.~Zhou, G.~Sun, and Y.~Wu.
\newblock Rigel: Transforming tabular data by declarative mapping.
\newblock {\em IEEE Transactions on Visualization and Computer Graphics}, 2022.

\bibitem{2013-Automatic-Chen}
Z.~Chen and M.~Cafarella.
\newblock Automatic web spreadsheet data extraction.
\newblock In {\em Proc. Int. Workshop on Semantic Search over the Web (SSW)},
  pp. 1:1--1:8, 2013.

\bibitem{2021-fuzzy-dhanoa}
V.~Dhanoa, C.~Walchshofer, A.~Hinterreiter, E.~Gr{\"o}ller, and M.~Streit.
\newblock Fuzzy spreadsheet: Understanding and exploring uncertainties in
  tabular calculations.
\newblock {\em IEEE Transactions on Visualization and Computer Graphics}, 2021.

\bibitem{2019-semantic-dong}
H.~Dong, S.~Liu, Z.~Fu, S.~Han, and D.~Zhang.
\newblock Semantic structure extraction for spreadsheet tables with a
  multi-task learning architecture.
\newblock In {\em Proc. Workshop on Document Intelligence}, 2019.

\bibitem{2018-Expandable-Dou}
W.~Dou, S.~Han, L.~Xu, D.~Zhang, and J.~Wei.
\newblock Expandable group identification in spreadsheets.
\newblock In {\em Proc. ACM/IEEE Int. Conf. Automated Software Engineering
  (ASE)}, pp. 498--508, 2018.

\bibitem{2016-venron-dou}
W.~Dou, L.~Xu, S.-C. Cheung, C.~Gao, J.~Wei, and T.~Huang.
\newblock Venron: A versioned spreadsheet corpus and related evolution
  analysis.
\newblock In {\em Proc. IEEE/ACM Int. Conf. Software Engineering Companion
  (ICSE-C)}, pp. 162--171, 2016.

\bibitem{2021-TabularNet-Du}
L.~Du, F.~Gao, X.~Chen, R.~Jia, J.~Wang, J.~Zhang, S.~Han, and D.~Zhang.
\newblock Tabularnet: {A} neural network architecture for understanding
  semantic structures of tabular data.
\newblock In {\em Proc. ACM Conf. Knowledge Discovery and Data Mining
  (SIGKDD)}, pp. 322--331, 2021.

\bibitem{2004-visualising-dwyer}
T.~Dwyer and D.~R. Gallagher.
\newblock Visualising changes in fund manager holdings in two and a
  half-dimensions.
\newblock {\em Information Visualisation}, 3(4):227--244, 2004.

\bibitem{2005-euses-fisher}
M.~Fisher and G.~Rothermel.
\newblock The euses spreadsheet corpus: A shared resource for supporting
  experimentation with spreadsheet dependability mechanisms.
\newblock In {\em Proc. Workshop on End-User Software Engineering (WEUSE)}, p.
  1–5, 2005.

\bibitem{2019-taggle-iv}
K.~Furmanova, S.~Gratzl, H.~Stitz, T.~Zichner, M.~Jaresova, A.~Lex, and
  M.~Streit.
\newblock Taggle: Scalable visualization of tabular data through aggregation.
\newblock {\em Information Visualisation}, 19(2):114--136, 2019.

\bibitem{2006-integrating-ginns}
P.~Ginns.
\newblock Integrating information: A meta-analysis of the spatial contiguity
  and temporal contiguity effects.
\newblock {\em Learning and Instruction}, 16(6):511--525, 2006.

\bibitem{2014-Domino-Gratzl}
S.~Gratzl, N.~Gehlenborg, A.~Lex, H.~Pfister, and M.~Streit.
\newblock Domino: Extracting, comparing, and manipulating subsets across
  multiple tabular datasets.
\newblock {\em IEEE Transactions on Visualization and Computer Graphics},
  20(12):2023--2032, 2014.

\bibitem{2013-Lineup-Gratzl}
S.~Gratzl, A.~Lex, N.~Gehlenborg, H.~Pfister, and M.~Streit.
\newblock Lineup: Visual analysis of multi-attribute rankings.
\newblock {\em IEEE Transactions on Visualization and Computer Graphics},
  19(12):2277--2286, 2013.

\bibitem{2016-heatmap-gu}
Z.~Gu, R.~Eils, and M.~Schlesner.
\newblock Complex heatmaps reveal patterns and correlations in multidimensional
  genomic data.
\newblock {\em Bioinformatics}, 32(18):2847--2849, 2016.

\bibitem{2012-Spreadsheet-Gulwani}
S.~Gulwani, W.~R. Harris, and R.~Singh.
\newblock Spreadsheet data manipulation using examples.
\newblock {\em Communications of the ACM}, 55(8):97--105, 2012.

\bibitem{2011-Spreadsheet-Harris}
W.~R. Harris and S.~Gulwani.
\newblock Spreadsheet table transformations from examples.
\newblock In {\em Proc. ACM SIGPLAN Conf. Programming Language Design and
  Implementation (PLDI)}, pp. 317--328, 2011.

\bibitem{2009-horizon}
J.~Heer, N.~Kong, and M.~Agrawala.
\newblock Sizing the horizon: The effects of chart size and layering on the
  graphical perception of time series visualizations.
\newblock In {\em Proc. ACM Conf. Human Factors in Computing Systems (CHI)},
  pp. 1303--1312, 2009.

\bibitem{2011-spreadsheet-hung}
V.~Hung, B.~Benatallah, and R.~Saint-Paul.
\newblock Spreadsheet-based complex data transformation.
\newblock In {\em Proc. ACM Conf. Information and Knowledge Management (CIKM)},
  pp. 1749--1754, 2011.

\bibitem{2011-Wrangler-Sean}
S.~Kandel, A.~Paepcke, J.~M. Hellerstein, and J.~Heer.
\newblock Wrangler: interactive visual specification of data transformation
  scripts.
\newblock In {\em Proc. ACM Conf. Human Factors in Computing Systems (CHI)},
  pp. 3363--3372, 2011.

\bibitem{2011-visbrick}
A.~Lex, H.-J. Schulz, M.~Streit, C.~Partl, and D.~Schmalstieg.
\newblock Visbricks: Multiform visualization of large, inhomogeneous data.
\newblock {\em IEEE Transactions on Visualization and Computer Graphics},
  17(12):2291--2300, 2011.

\bibitem{2020-gotree-li}
G.~Li, M.~Tian, Q.~Xu, M.~J. McGuffin, and X.~Yuan.
\newblock Go{T}ree: A grammar of tree visualizations.
\newblock In {\em Proc. ACM Conf. Human Factors in Computing Systems (CHI)},
  pp. 1--13, 2020.

\bibitem{2020-barcodetree-li}
G.~Li, Y.~Zhang, Y.~Dong, J.~Liang, J.~Zhang, J.~Wang, M.~J. Mcguffin, and
  X.~Yuan.
\newblock Barcode{T}ree: Scalable comparison of multiple hierarchies.
\newblock {\em IEEE Transactions on Visualization and Computer Graphics},
  26(1):1022--1032, 2020.

\bibitem{2018-TACO-Niederer}
C.~Niederer, H.~Stitz, R.~Hourieh, F.~Grassinger, W.~Aigner, and M.~Streit.
\newblock Taco: Visualizing changes in tables over time.
\newblock {\em IEEE Transactions on Visualization and Computer Graphics},
  24(1):677--686, 2018.

\bibitem{2006-AyersSweller}
A.~Paul and S.~John.
\newblock The split-attention principle in multimedia learning.
\newblock {\em The Cambridge Handbook of Multimedia Learning}, 2:135--146,
  2005.

\bibitem{2014-Revisiting-charles}
C.~Perin, P.~Dragicevic, and J.~Fekete.
\newblock Revisiting bertin matrices: New interactions for crafting tabular
  visualizations.
\newblock {\em IEEE Transactions on Visualization and Computer Graphics},
  20(12):2082--2091, 2014.

\bibitem{2001-PotterWheel-Raman}
V.~Raman and J.~M. Hellerstein.
\newblock Potter's wheel: An interactive data cleaning system.
\newblock In {\em Proc. Int. Conf. Very Large Data Bases (VLDB)}, pp. 381--390,
  2001.

\bibitem{1994-table-rao}
R.~Rao and S.~K. Card.
\newblock The table lens: merging graphical and symbolic representations in an
  interactive focus + context visualization for tabular information.
\newblock In {\em Proc. ACM Conf. Human Factors in Computing Systems (CHI)},
  pp. 318--322, 1994.

\bibitem{2015-table-shigarov}
A.~O. Shigarov.
\newblock Table understanding using a rule engine.
\newblock {\em Expert Systems with Applications}, 42(2):929--937, 2015.

\bibitem{2019-TabbyXL-Alexey}
A.~O. Shigarov, V.~V. Khristyuk, A.~A. Mikhailov, and V.~Paramonov.
\newblock Tabbyxl: Rule-based spreadsheet data extraction and transformation.
\newblock In {\em Proc. Int. Conf. Information and Software Technologies
  (ICIST)}, vol. 1078, pp. 59--75, 2019.

\bibitem{1996-eyes-ben}
B.~Shneiderman.
\newblock The eyes have it: a task by data type taxonomy for information
  visualizations.
\newblock In {\em Proc. IEEE Symp. Visual Languages (VL)}, pp. 336--343, 1996.

\bibitem{2005-constructing-harri}
H.~Siirtola and E.~Makinen.
\newblock Constructing and reconstructing the reorderable matrix.
\newblock {\em Information Visualisation}, 4(1):32--48, 2005.

\bibitem{1996-focus-spenke}
M.~Spenke, C.~Beilken, and T.~Berlage.
\newblock Focus: the interactive table for product comparison and selection.
\newblock In {\em Proc. ACM Conf. Symposium on User Interface Software and
  Technology (UIST)}, pp. 41--50, 1996.

\bibitem{2017-Transforming-Su}
H.~Su, Y.~Li, X.~Wang, G.~Hao, Y.~Lai, and W.~Wang.
\newblock Transforming a nonstandard table into formalized tables.
\newblock In {\em Proc. IEEE Conf. Web Information Systems and Applications
  (WISA)}, pp. 311--316, 2017.

\bibitem{2005-implications-sweller}
J.~Sweller.
\newblock Implications of cognitive load theory for multimedia learning.
\newblock {\em The Cambridge Handbook of Multimedia Learning}, 3(2):19--30,
  2005.

\bibitem{1998-cognitive-sweller}
J.~Sweller, J.~J. Van~Merrienboer, and F.~G. Paas.
\newblock Cognitive architecture and instructional design.
\newblock {\em Educational Psychology Review}, 10(3):251--296, 1998.

\bibitem{2022-Simultaneous-Beusekom}
N.~van Beusekom, W.~Meulemans, and B.~Speckmann.
\newblock Simultaneous matrix orderings for graph collections.
\newblock {\em IEEE Transactions on Visualization and Computer Graphics},
  28(1):1--10, 2022.

\bibitem{2017-podium-wall}
E.~Wall, S.~Das, R.~Chawla, B.~Kalidindi, E.~T. Brown, and A.~Endert.
\newblock Podium: Ranking data using mixed-initiative visual analytics.
\newblock {\em IEEE Transactions on Visualization and Computer Graphics},
  24(1):288--297, 2017.

\bibitem{1996-xtable-wang}
X.~Wang and D.~Wood.
\newblock Xtable: a tabular editor and formatter.
\newblock Technical report, 1996.
\newblock
  \url{http://cajun.cs.nott.ac.uk/compsci/epo/papers/volume8/issue2/2point6.pdf}.

\bibitem{2018-BiDots-Zhao}
J.~Zhao, M.~Sun, F.~Chen, and P.~Chiu.
\newblock Bi{D}ots: Visual exploration of weighted biclusters.
\newblock {\em IEEE Transactions on Visualization and Computer Graphics},
  24(1):195--204, 2018.

\end{thebibliography}
\end{document}